\documentclass{article}

\usepackage[preprint,nonatbib]{neurips_2023}

\usepackage[utf8]{inputenc} 
\usepackage[T1]{fontenc}    
\usepackage{hyperref}       
\usepackage{url}            
\usepackage{booktabs}       
\usepackage{amsfonts}       
\usepackage{nicefrac}       
\usepackage{microtype}      
\usepackage{xcolor}         
\usepackage{color}
\usepackage{wrapfig}
\usepackage{tikz}
\usetikzlibrary{tikzmark,decorations.pathreplacing,calligraphy}

\usepackage{amsmath,amsthm,amssymb,amsfonts}
\usepackage{bm}
\usepackage{multirow}
\usepackage{algorithm}
\usepackage{algpseudocode}
\newtheorem{lemma}{Lemma}
\newtheorem{theorem}{Theorem}
\newtheorem{definition}{Definition}

\title{Re-thinking Data Availablity Attacks Against Deep Neural Networks}

\begin{document}
\author{Bin Fang$^{1*}$\quad  Bo Li$^{2*}$\quad Shuang Wu$^{2}$\quad \\ \textbf{Ran Yi}$^{1}$\quad  \textbf{Shouhong Ding}$^{2}$\quad
\textbf{Lizhuang Ma}$^{1}$
\vspace{0.2cm}\\ 
\vspace{0.2cm}
$^1$Shanghai Jiao Tong University \qquad $^2$Youtu Lab Tencent \\
}

\maketitle
\renewcommand{\thefootnote}{\fnsymbol{footnote}}
\footnotetext[1]{Both authors contributed equally to this work. Work done during Bin Fang's internship at Tencent Youtu Lab and Bo Li is the project lead.}

\begin{abstract}
The unauthorized use of personal data for commercial purposes and the clandestine acquisition of private data for training machine learning models continue to raise concerns. In response to these issues, researchers have proposed \textit{availability attacks} that aim to render data unexploitable. However, many current attack methods are rendered ineffective by adversarial training. In this paper, we re-examine the concept of unlearnable examples and discern that the existing robust error-minimizing noise presents an inaccurate optimization objective. Building on these observations, we introduce a novel optimization paradigm that yields improved protection results with reduced computational time requirements. We have conducted extensive experiments to substantiate the soundness of our approach. Moreover, our method establishes a robust foundation for future research in this area.
\end{abstract}

\section{Introduction}

Over the last decade, remarkable advancements have been made in the field of Artificial Intelligence (AI), leading to significant impacts across a wide range of domains. The key driving force behind these impressive achievements has been the access to vast quantities of high-quality data. In fact, many major AI breakthroughs have been realized only after obtaining the appropriate training data. The recent advances in large foundation models~\cite{DBLP:conf/naacl/DevlinCLT19,DBLP:journals/corr/abs-2303-08774} and generative models~\cite{DBLP:conf/cvpr/RombachBLEO22,DBLP:conf/mindtrek/Oppenlaender22a} stand as strong evidence. 
Nonetheless, behind these remarkable accomplishments lies an issue that cannot be overlooked: the unauthorized collection and utilization of data. There is evidence to suggest that technology corporations are engaged in the collection and utilization of unauthorized data for the purpose of training their commercial models~\cite{openai,midjounary&sd,midjounary1}.


To mitigate the unauthorized use of data, availability attacks have been proposed~\cite{Availability-Attack}. Numerous studies demonstrate that injecting imperceptible noise into data can considerably impair the performance of models reliant on such contaminated data~\cite{AA-1, TAP2021, NTGA2021, EM2021, ShortCuts2022, CUDA2023, UC2023}. Nevertheless, the noise produced by these approaches can be readily neutralized by adversarial training, thereby undermining the protective efficacy with respect to the poisoned data. To address this limitation, Robust Error-Minimizing (REM) noise~\cite{REM2022} has been introduced as a means to diminish the detrimental impact of adversarial training on unlearnable examples.

Although REM noise partially attenuates the detrimental effects of adversarial training on unlearnable examples, its theoretical underpinnings remain dubious. Upon scrutinizing its optimization objective, we observe that this objective closely resembles that of Error-Minimizing (EM)~\cite{EM2021}. Consequently, the efficacy of REM can be ascribed to the implementation of adversarial training in the underlying code, rather than the generation of adversarial examples, as initially claimed.

In this paper, we introduce a novel optimization process and objectives that inherently capture the essence of unlearnable examples, offering considerable protection against adversarial training. We posit that error-minimizing noise is insufficiently equipped to defend against adversarial training, as its generator training strategy relies on standard training, which extracts only non-robust features~\cite{ST2019}. Regarding robust error-minimizing noise~\cite{REM2022}, we contend that its undisclosed adversarial training procedure, rather than the adversarial examples elaborated upon in their work, accounts for its effectiveness.

Building on these observations, we propose that an unlearnable noise generator, trained using adversarial training techniques, can produce more robust protective noise to counter the destructive effects of adversarial training. Additionally, for the first time, we derive the intended effect of unlearnable noise according to its definition—a critical aspect overlooked by previous research. Subsequently, we modify our optimization objective to reflect this understanding. Owing to these key insights, our method achieves remarkable protective performance in both standard and adversarial training contexts. Furthermore, due to our enhanced optimization procedure, our method surpasses the efficiency of REM~\cite{REM2022}, rendering it suitable for deployment with large-scale models and datasets.

In summary, our key contributions are as follows:
\begin{itemize}
    \item We elucidate the inherent limitations of previous methods for generating unlearnable noise, thereby identifying areas for improvement.
    \item We put forth a sound optimization procedure that effectively attenuates the deleterious impact of adversarial training, providing robust protection against unauthorized data usage.
    \item For the first time, we formulate the intended effect of unlearnable examples and draw from this formulation to modify our optimization objective, which subsequently yields remarkable performance enhancements.
    \item We establish a compelling baseline for future research, enabling the facile incorporation of supplementary constraints into our optimization objective, thereby facilitating further advancements in this domain.
\end{itemize}

\section{Related Work}
\subsection{Poisoning Attacks}
Data poisoning attacks endeavor to undermine the training process of a model by infusing noise into the training dataset, leading to substantial testing errors on specific or unseen samples during the testing phase. Backdoor attacks constitute a prevalent form of data poisoning attack, often characterized by the injection of triggers into training samples, which subsequently provokes the misclassification of images containing these triggers during the testing phase~\cite{backdoor2021-1, backdoor2020-2, backdoor2020-3}. Nevertheless, it is important to note that such attacks typically influence only samples with trigger patterns, while leaving clean samples unaffected and correctly classified~\cite{backdoor2017-5, backdoor2018-4}.


\subsection{Availability Attacks}
Availability Attacks aim to safeguard data from unauthorized exploitation by generating imperceptible, unlearnable noise. The data compromised by this type of attack are referred to as unlearnable examples. Deep neural networks trained on unlearnable examples exhibit performance analogous to random guessing on standard test examples. 

\subsubsection{Model-free Attacks}
These attacks typically generate unlearnable noise directly from the pixel level, rather than the feature level. Consequently, methods in this category, such as LSP~\cite{ShortCuts2022} and CUDA~\cite{CUDA2023}, do not require any feature information on clean data, yielding an irrelevant relationship with data features. Owing to their pixel-level approach, these methods are highly efficient. However, the intrinsic design principle of this approach renders unlearnable examples vulnerable to feature-based defense methods, such as adversarial training—a fundamental flaw that cannot be circumvented.

\subsubsection{Model-based Attacks}
These attacks usually generate unlearnable noise through surrogate models. This category of methods trains a surrogate model, also known as a noise generator, and learns data features while simulating the training process of the poisoned model during the surrogate model's training phase.

\textbf{Non-robust Model-based Attacks.}
This type of attack involves training surrogate models as non-robust models that learn non-robust features, such as TAP~\cite{TAP2021}, NTGA~\cite{NTGA2021}, and EM~\cite{EM2021}. Consequently, the unlearnable noise generated by this approach only targets poisoned models subjected to standard training and merely prevents models from learning standard data features. Once the poisoned models undergo adversarial training, the protective effects of these methods are disrupted.

\textbf{Robust Model-based Attacks.} 
This type of attack entails training surrogate models as robust models that learn robust features. REM~\cite{REM2022} is the only representative work in this category. REM~\cite{REM2022} posits that poisoned models undergoing adversarial training learn knowledge from adversarial examples. As a result, REM~\cite{REM2022} generates unlearnable noise exclusively for adversarial examples, rather than clean data, ensuring that unlearnable noise remains effective against adversarial training. Let $\mathcal{D} = \{(x_1,y_1),(x_2,y_2),\dots,(x_n,y_n)\}$ represent a dataset consisting of $n$ samples, where $x_i \in \mathcal{X}$ is the $i$-th sample and $y_i \in \mathcal{Y} = \{1,\dots, K\}$ is the corresponding label. Let a parameterized machine learning model be denoted by $f_{\theta}: \mathcal{X}\rightarrow \mathcal{Y}$, where $\theta \in \Omega$ is the model parameter. Let $\ell$ denote a loss function. Then, the training objective for the robust noise generator is as follows:
\begin{equation}
    \min_{\theta} \frac{1}{n} \sum_{i=1}^n \min_{\|\delta^u_i\|\leq \rho_u} \max_{\|\delta^a_i\| \leq \rho_a} \ell(f'_\theta(x_i+\delta^u_i+\delta^a_i),y_i)
\end{equation} 
where $\rho_u$ is the defensive perturbation radius that compels the generated robust unlearnable noise to be imperceptible.

Upon examining the training objective, it becomes evident that the trained surrogate model is also a standardly trained model, which implies that it captures only non-robust features, rather than robust features~\cite{ST2019}. As previously discussed, this type of unlearnable noise cannot protect data from adversarial training. However, REM~\cite{REM2022} remains effective due to the adversarial training procedure in their implementation, an aspect not addressed in their theoretical framework.

\section{Methodology}
Based on our previous analysis, we posit that a robust surrogate model is crucial for generating unlearnable noise that can effectively protect data during adversarial training. In Section~\ref{sec:optimization_procedure}, we propose a two-stage optimization procedure for training a robust surrogate model, addressing the shortcomings of existing robust model-based availability attacks. Furthermore, recognizing that current availability attacks overlook the definition of the effects of models trained on unlearnable examples, in Section~\ref{sec:optimization_object}, we initially formulate the effects of models trained on unlearnable examples and subsequently modify the optimization objective for training a more effective robust surrogate model.

\subsection{Two-Stage Optimization Procedure}
\label{sec:optimization_procedure}
In light of our analysis, we propose a two-stage min-max-min optimization process to train a robust surrogate model capable of generating robust, unlearnable noise. The first stage consists of an inner minimization process, where unlearnable noise is obtained for a noise generator that has undergone adversarial training. Since adversarial training can extract robust features, the unlearnable noise generated by robust models can naturally resist adversarial training. The second stage is an external min-max optimization process equivalent to adversarial training. The input for this stage comprises images with robust unlearnable noise added, such that the external procedure simulates the adversarial training process and closely resembles the training process of a poisoned model using adversarial training. Consequently, both the first and second stages complement each other, and the internal generation of unlearnable noise yields better protective effects on the robust model.

The two-stage min-max-min optimization process is specified below:
\begin{enumerate}
    \item[a)] Generate unlearnable examples $x'$ from the surrogate model $f'_{\theta}$ by \begin{equation}
        \delta_u = \min_{||\delta_i^u||\leq \rho_u} \ell(f'_{\theta}(x_i+\delta_i^u),y_i)
        \label{stage-one}
    \end{equation}
    \item[b)] Perform adversarial training of the surrogate model $f'_{\theta}$ to extract the robust features of the unlearnable examples
    \begin{equation}
        \min_{\theta} \frac{1}{n}\sum_{i=1}^{n} \max_{||\delta_i^a|| \leq \rho_a}\ell(f'_{\theta}(x_i+\delta_i^u+\delta_i^a),y_i)
        \label{stage-two}
    \end{equation}
\end{enumerate}
Additionally, we expect $ \rho_a \leq \rho_u $.

\subsection{Optimization Objective}
\label{sec:optimization_object}
EM~\cite{EM2021} observes that a model trained on unlearnable examples should exhibit random guessing behavior on clean samples. Noting that most previous optimization objectives did not constrain performance on clean samples, we first formulate randomness and use it to modify our optimization objective.
The formulation is as follows:

\begin{definition}[Averaged Prediction Randomness]
Let $\mathcal{D} = \{(x_1,y_1),(x_2,y_2),\dots,(x_N,y_N)\}$ indicate a dataset consisting of $N$ samples, where $x_i \in \mathcal{X}$ is the $i$-th sample and $y_i \in \mathcal{Y} = \{1,\dots, K\}$ is the corresponding label. Let a classifier denote $C: \mathcal{X}\rightarrow \mathcal{Y}$. Let $P_k$ be the probability vector of model predictions on samples with ground-truth label $k$, where the $j$-th element of $P_k$ is  
\begin{equation}
    P_k^j \triangleq \frac{\sum_{i=1}^N \mathbb{I}\{C(x_i)=j\} \cdot \mathbb{I}\{y_i=k\}}{\sum_{i=1}^N  \mathbb{I}\{y_i=k\}} 
\end{equation}
The average prediction randomness metric $R_p$ is defined as 
\begin{equation}
    R_p \triangleq \frac{1}{N} \sum_{k=1}^K \sum_{i=1}^N \mathbb{I}\{y_i=k\} \cdot \mathcal{L}\left({P}_{k}\right)
\end{equation}
Where $\mathcal{L} (\cdot)$ denotes a distance function. 
\end{definition}
In general, $R_p$ measures the distance between the current predicted distribution and a uniform distribution. The smaller the value of $R_p$, the better the dispersion. However, $R_p$ is non-differentiable and cannot be optimized directly. We introduce a differentiable modified formulation to alleviate this problem, as follows:

\begin{definition}[Averaged Sample-wise Randomness]
Let $\mathcal{D} = \{(x_1,y_1),(x_2,y_2),\dots,(x_N,y_N)\}$ represent a dataset consisting of $N$ samples, where $x_i \in \mathcal{X}$ is the $i$-th sample and $y_i \in \mathcal{Y} = \{1,\dots, K\}$ is the corresponding label. Let a classifier be denoted as $C: \mathcal{X}\rightarrow \mathcal{Y}$. Let a parameterized machine learning model be represented as $f_{\theta}$, where $\theta \in \Omega$ is the model parameter. The averaged sample-wise randomness of predictions given by the classifier $f_{\theta}(\cdot)$ is defined as 
\begin{equation}
    R_s \triangleq \frac{1}{N}\sum_{i=1}^{N}  \mathcal{L}\left(f({x}_i)\right)
\end{equation}
\end{definition}
In general, the averaged sample-wise randomness characterizes the average dispersion of predicted probability vectors for all samples.

Metrics that measure the distance between the predicted distribution and a uniform distribution are of vital importance. Cross-entropy and KL divergence are commonly employed to measure the distance between distributions; however, neither constitutes a distance function. Consequently, we opt for mean square error, a widely used distance measurement. Additionally, SCORE~\cite{Robustness2022} observes that models trained with a distance loss function frequently outperform those trained with a non-distance loss function. We have demonstrated that cross-entropy loss and KL loss are equivalent for assessing the distance between the predicted distributions of $f_{\theta}$ and uniform distributions. Furthermore, under our setting, we establish that KL loss constitutes an upper bound of mean square error loss. For KL loss, when the prediction probability of a specific class $f_\theta(\cdot)[k]$ is exceedingly small, the loss becomes infinite, as does the gradient, leading to gradient explosion and complicating the training process. Conversely, mean square error exhibits relative smoothness within its value range—with well-defined upper and lower bounds—thereby facilitating model training. As a result, we select mean square error as the loss metric to gauge sample-wise randomness. Consequently, the smaller the ASR, the more random the output probability of $f_{\theta}$ on clean samples becomes, revealing the extent of knowledge acquired by the model.

\begin{wrapfigure}{r}{8.4cm}
        \centering
        \vspace{-0.2cm}
        \includegraphics[scale=0.1]{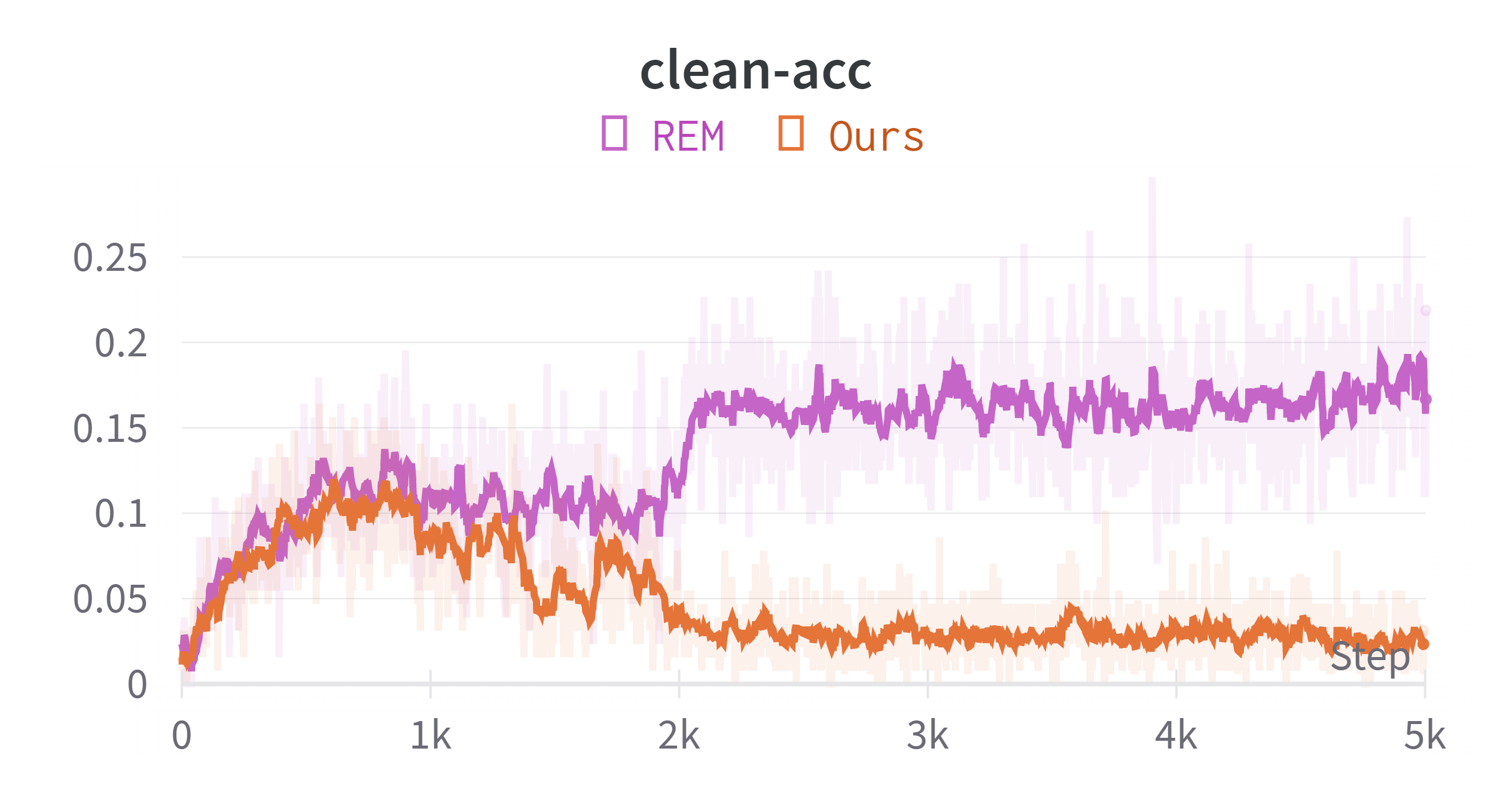}
        \vspace{-0.5cm}
        \caption{Test accuracies of clean training examples.}
        \label{fig:clean-acc}
 \vspace{-0.3cm}
\end{wrapfigure}




\begin{lemma}
Let $f_{\theta}(x_i)[k] $ indicates the $k$-th value of the predicted vector. Let $\mathcal{R}=(\frac{1}{K},\dots, \frac{1}{K})\in R^K$ denotes the random guess probability and $\Theta = (f_{\theta}(x_i)[0], \dots, f_{\theta}(x_i)[K]) \in R^K$ denotes the predicted probability. Then, we have 
\begin{equation}\nonumber
     -\mathcal{R}\log\Theta = \text{KL}(\mathcal{R}|\Theta)  + \log K.
\end{equation}
\end{lemma}
It has been proven that cross-entropy loss is equivalent to KL loss in measuring the distance between the predicted probability and a random guess distribution.
It is clear that $-\mathcal{R}\log\Theta \geq \log K$.

\begin{theorem}
Let $f_{\theta}(x_i)[k] $ indicates the $k$-th value of the predicted vector. Let $\mathcal{R}=(\frac{1}{K},\dots, \frac{1}{K})\in R^K$ denotes the random guess probability. Then, we have
\begin{equation}\nonumber
    0 \leq \frac{1}{K}\sum_{j=1}^K(f_{\theta}(x)[j]-\frac{1}{K})^2 \leq \frac{4}{K}
\end{equation}
\end{theorem}
If and only if $f_{\theta}(x)[i] = f_{\theta}(x)[j], i\neq j$, equation sign is established.

In most cases (under our experimental settings), the mean square error is smaller than the cross-entropy loss. Moreover, both loss functions achieve their minimum values under the same conditions, indicating that optimizing mean square error is equivalent to optimizing cross-entropy. Furthermore, squared error loss exhibits smoothness and does not suffer from a gradient explosion problem, making it a superior optimization objective.

In summary, the modified optimization objective of the second step is as follows:
\begin{equation}
    \min_{\theta} \frac{1}{n}\sum_{i=1}^{n} \bigg [ \max_{||\delta_i^u|| \leq \rho_a}\ell( f'_{\theta}(x_i+\delta_i^u+\delta_i^a),y_i) + \frac{1}{K}\sum_{k=1}^{K} \bigg ( f'_{\theta}(x_i)[k]-\frac{1}{K}\bigg)^2 \bigg ] 
    \label{stage-two-modify}
\end{equation}

Further details are provided in Algorithm~\ref{algo:m1_noise_gen}. Figure~\ref{fig:clean-acc} shows the test accuracies of training examples when training the noise generator. It can be seen that the noise generated by REM can still make the surrogate learn a lot of information. Once we use our proposed ASR to modify the optimization object, the noise generated by our method can make the surrogate almost learn nothing. This reflects that our method generates stronger unlearnable noise.

\begin{algorithm}[h]
\caption{Training the noise generator}
\label{algo:m1_noise_gen}
\begin{algorithmic}[1]
\Require
Training data set $\mathcal{D}$,
training iteration $M$, 
classes $K$, 
learning rate $\eta$\\
PGD parameters $\rho_u$, $\alpha_u$ and $K_u$ for stage $1$, \\
PGD parameters $\rho_a$, $\alpha_a$ and $K_a$ for stage $2$. 
\Ensure Our noise generator $f'_\theta$.
    \State Initialize source model parameter $\theta$.
    \For{$i$ \textbf{in} $1, \cdots, M$} 
        \State Sample a minibatch $(x, y) \sim \mathcal{D}$.
        \State Initialize $\delta^u$.
        \For{$k$ \textbf{in} $1,\cdots, K_u$}   \tikzmark{a}
            \State $g_k \leftarrow  \frac{\partial}{\partial \delta^u} \ell(f'_\theta(x+\delta^u), y)$
            \State $\delta^u \leftarrow \prod_{\|\delta\|\leq\rho_u} \left( \delta^u - \alpha_u \cdot \mathrm{sign}(g_k) \right)$
        \EndFor \tikzmark{b} 
        \For{$k$ \textbf{in} $1,\cdots, K_a$} \tikzmark{c} 
            \State $g_k \leftarrow  \frac{\partial}{\partial \delta^u} \ell(f'_\theta(x+\delta^u+\delta^a), y)$
            \State $\delta^a \leftarrow \prod_{\|\delta\|\leq\rho_a} \left( \delta^a - \alpha_a \cdot \mathrm{sign}(g_k) \right)$
        \EndFor 
        \State $g_k \leftarrow \frac{\partial}{\partial \theta} [ \ell(f'_\theta(x+\delta^u+\delta^a), y) + \frac{1}{K}\sum_{k=1}^{K} ( f'_{\theta}(x_i)[k]-\frac{1}{K})^2 ]$ 
        \State $\theta \leftarrow \theta - \eta \cdot g_k$ \tikzmark{d}
    \EndFor
    \State \Return $f'_\theta$
\begin{tikzpicture}[remember picture,overlay]
  \path (pic cs:a) -|node(test){} (pic cs:b);
  \draw[overlay,decorate,
  decoration={calligraphic brace,amplitude=1mm},
  ultra thick] ([xshift=25em,yshift=0.05em]test.center) -- 
  ([xshift=25em]pic cs:b) 
  node[xshift=5mm,anchor=west,midway]
  {stage1};
\end{tikzpicture}
\begin{tikzpicture}[remember picture,overlay]
  \path (pic cs:c) -|node(test){} (pic cs:d);
  \draw[overlay,decorate,
  decoration={calligraphic brace,amplitude=1mm},
  ultra thick] ([xshift=21.85em,yshift=0.05em]test.center) -- 
  ([xshift=21.85em]pic cs:d) 
  node[xshift=3mm,anchor=west,midway]
  {stage2};
\end{tikzpicture}

\end{algorithmic}
\end{algorithm}

\section{Experiments}
In this section, we have conducted sufficient experiments to demonstrate the effectiveness of our method from various aspects. 
What's more, our method also presents amazing performances in generalization ability. The detailed experiments are as follows.

\subsection{Experiment Setup}
\textbf{Datasets.} 
To verify the effectiveness of our method on images of different categories and sizes, three commonly used datasets, namely CIFAR-10, CIFAR-100~\cite{CIFAR10}, and ImageNet subset~\cite{ImageNet2015} (consists of the first 100 classes), are used in our experiments. The data augmentation technique~\cite{DataAugment2019} is adopted in every experiment.

\textbf{Surrogate Models.}
Following~\cite{EM2021} and~\cite{REM2022}, we employ ResNet-18\cite{Resnet18} as the surrogate model $f'_{\theta}$ for trainging our noise generator with Eq.~(\ref{stage-one}) and Eq.~(\ref{stage-two-modify}). The $L_\infty$-bounded noises $\|\delta_u\|_\infty \leq \rho_u$ is adopted in our experiments. Additionally, we also use other surrogate models, including VGG-16\cite{VGG11}, ResNet-50\cite{Resnet18}, and DenseNet-121\cite{Densenet121}, to test the generalization ability of our method.

\textbf{Compared Methods.}
Our proposed method is compared with other state-of-the-art availability attacks, \textbf{TAP}~\cite{TAP2021}, \textbf{NTGA}~\cite{NTGA2021}, \textbf{EM}~\cite{EM2021}, and \textbf{REM}~\cite{REM2022}. 

\textbf{Noise Test.}
Noise generated by our method is tested both on standard training and adversarial training~\cite{AdversarialTraining2018}. We focus on $L_\infty$-bounded noise $\|\rho_a\|_\infty \leq \rho_a$ in adversarial training. We conduct adversarial training on unlearnable examples created by our method with different models, including VGG-16~\cite{VGG11}, ResNet-18, ResNet-50~\cite{Resnet18}, DenseNet-121~\cite{Densenet121}, and wide ResNet-34-10~\cite{WRN2016}. 
Note that when $\rho_a$ takes $0$, the adversarial training degenerates to the standard training.

\textbf{Metric.}
We evaluate the data protection ability of defensive noise by measuring the test accuracy of the model. A low test accuracy indicates that the model has learned little from the unlearnable examples, implying a strong protection ability from the noise.

\subsection{Significant Effectiveness of Our Method}
\subsubsection{Different adversarial training perturbation radii.}
To assess the robustness against adversarial training, we first introduce unlearnable noise to the entire training set, generating unlearnable examples. The unlearnable noise perturbation radius, denoted as $\rho_u$, is set to $8/255$ for all noise-generating methods and the adversarial perturbation radius $\rho_a$ is set as $4/255$ for REM~\cite{REM2022} and our method. Subsequently, we train models using different adversarial training perturbation radii $\rho_a$ on these unlearnable examples. It is important to note that when $\rho_a=0$, the models are trained using the standard training method.
Table~\ref{tab:diff_radius} reports the accuracies of the trained models on the unlearnable examples.

\begin{table}[h]
\centering
\caption{Test accuracy (\%) of models trained on data protected by different availability attacks via adversarial training with different perturbation radii.}
\label{tab:diff_radius}
\scriptsize
\begin{tabular}{c c c c c c c c}
\toprule
\multirow{2}{1.2cm}{\centering Dataset} & \multirow{3}{1.2cm}{\centering Adv. Train. \\ $\rho_a$} & \multirow{2}{0.8cm}{\centering Clean} & \multirow{2}{0.8cm}{\centering EM} & \multirow{2}{0.8cm}{\centering TAP} & \multirow{2}{0.8cm}{\centering NTGA} & {\centering REM} & {\centering Ours} \\
\cmidrule(lr){7-8}
& & & & & & $\rho_a = 4/255$ & $\rho_a = 4/255$ \\
\midrule
\multirow{5}{1.2cm}{\centering CIFAR-10}
& $0$     & 94.66 & 13.20 & 22.51 & 16.27 & 22.93 & \textbf{12.71} \\
& $1/255$ & 93.74 & 22.08 & 92.16 & 41.53 & 30.00 & \textbf{14.71}  \\
& $2/255$ & 92.37 & 71.43 & 90.53 & 85.13 & 30.04 & \textbf{15.38} \\
& $3/255$ & 90.90 & 87.71 & 89.55 & 89.41 & 31.75 & \textbf{15.51} \\
& $4/255$ & 89.51 & 88.62 & 88.02 & 88.96 & 48.16 & \textbf{23.12} \\
\midrule
\multirow{5}{1.2cm}{\centering CIFAR-100}
& $0$     & 76.27 &  \textbf{1.60} & 13.75 & 3.22 & 11.63 & 3.27  \\
& $1/255$ & 71.90 & 71.47 & 70.03 & 65.74 & 14.48 & \textbf{7.79}  \\ 
& $2/255$ & 68.91 & 68.49 & 66.91 & 66.53 & 16.60 & \textbf{7.73}  \\ 
& $3/255$ & 66.45 & 65.66 & 64.30 & 64.80 & 20.70 & \textbf{9.91}  \\ 
& $4/255$ & 64.50 & 63.43 & 62.39 & 62.44 & 27.35 & \textbf{23.00} \\ 
\midrule
\multirow{5}{1.2cm}{\centering ImageNet Subset}
& $0$     & 80.66 &  \textbf{1.26} & 9.10 & 8.42 & 13.74 & 4.08 \\
& $1/255$ & 76.20 & 74.88 & 75.14 & 63.28 & 21.58 & \textbf{11.80} \\
& $2/255$ & 72.52 & 71.74 & 70.56 & 66.96 & 29.40 & \textbf{16.88} \\
& $3/255$ & 69.68 & 66.90 & 67.64 & 65.98 & 35.76 & \textbf{22.34} \\
& $4/255$ & 66.62 & 63.40 & 63.56 & 63.06 & 41.66 & \textbf{31.64} \\
\bottomrule
\end{tabular}
\vspace{-4mm}
\end{table}

As shown in Table~\ref{tab:diff_radius}, the adversarial training perturbation $\rho_a$ ranges from $1/255$ to $4/255$. The surrogate models are ResNet-18~\cite{Resnet18}. For adversarial training, we can find that even a very small adversarial training perturbation radius of $2/255$ can damage the protecting effects of TAP~\cite{TAP2021}, NTGA~\cite{NTGA2021}, and EM\cite{EM2021}. 
Table~\ref{tab:diff_radius} illustrates that when the unlearnable noise perturbation radius is fixed, the protective effect diminishes as the adversarial training perturbation radius increases. This observation suggests that to safeguard data against adversarial training with a perturbation radius $\rho_a$, one must set the unlearnable perturbation radius $\rho_u$ of robust methods to a value relatively larger than $\rho_a$. Notably, our method consistently outperforms other approaches by a significant margin, regardless of the adversarial perturbation radii. In standard training scenarios, our method also exhibits superior performance compared to most alternatives.

Moreover, our method maintains substantial protective effects across different datasets, regardless of their size or class composition, especially when subjected to adversarial training. Overall, these experiments demonstrate that our method is capable of effectively protecting data across various datasets and adversarial training perturbation radii.

\subsubsection{Different protection percentages.}
Realistically, there is a more challenging scenario, where only a part of the data is protected by the unlearnable noise, while the others are clean.
Specifically, we randomly select a part of the training data from the whole training set, adding unlearnable noise to the selected data. We then conduct adversarial training with ResNet-18 on the mixed data and the remaining clean data.
The unlearnable perturbation radius for every noise is set as $8/255$, while the adversarial perturbation radius $\rho_a$ of REM and our noise is set as $4/255$.
The difference between the test accuracies on mixed data and clean data reflects the knowledge gained from the protected training data.
The accuracies on clean test data are reported in Table~\ref{tab:percentage}.

\begin{table}[ht]
\centering
\caption{Test accuracy (\%) on CIFAR-10 and CIFAR-100 with different protection percentages.}
\label{tab:percentage}
\scriptsize
\begin{tabular}{c c c c | c c | c c | c c | c c | c}
\toprule
\multirow{4}{1.2cm}{\centering Dataset} &
\multirow{4}{0.6cm}{\centering Adv. \\ Train. \\ $\rho_a$} & \multirow{3}{0.6cm}{\centering Noise \\ Type} & \multicolumn{10}{c}{\centering Data Protection Percentage} \\
\cmidrule{4-13}
&& & \multirow{2}{0.4cm}{\centering 0\%} & \multicolumn{2}{c|}{\centering 20\%} & \multicolumn{2}{c|}{\centering 40\%} & \multicolumn{2}{c|}{\centering 60\%} & \multicolumn{2}{c|}{\centering 80\%} & \multirow{2}{0.4cm}{\centering 100\%} \\
& & & & Mixed & Clean & Mixed & Clean & Mixed & Clean & Mixed & Clean & \\
\midrule
\multirow{11}{1.2cm}{\centering CIFAR-10} &
\multirow{5}{0.5cm}{\centering $2/255$} & EM & \multirow{5}{0.5cm}{\centering 92.37} & 92.33 & \multirow{5}{0.5cm}{\centering 91.30} & 92.18 & \multirow{5}{0.5cm}{\centering 90.31} & 92.00 & \multirow{5}{0.5cm}{\centering 88.65} & 92.06 & \multirow{5}{0.5cm}{\centering 83.37} & 71.43 \\
&& TAP &
& 92.17 && 91.62 && 91.32 && 91.48 && 90.53 \\
&& NTGA &
& 92.41 && 92.19 && 92.23 && 91.74 && 85.13 \\
&& REM &
& 92.23 && 90.79 && 88.85 && 83.70 && 30.04 \\
&& Ours &
& \textbf{92.03} && \textbf{90.34} && \textbf{87.98} && \textbf{83.32} && \textbf{15.38} \\
\cmidrule{2-13}
& \multirow{5}{0.5cm}{\centering $4/255$} & EM & \multirow{5}{0.5cm}{\centering 89.51} & 89.39 & \multirow{5}{0.5cm}{\centering 88.17} & 89.09 & \multirow{5}{0.5cm}{\centering 86.76} & 89.41 & \multirow{5}{0.5cm}{\centering 85.07} & 89.41 & \multirow{5}{0.5cm}{\centering 79.41} & 88.62 \\
&& TAP &
& 89.01 && 88.66 && 88.40 && 88.04 && 88.02 \\
&& NTGA &
& 89.56 && 89.35 && 89.22 && 89.17 && 88.96 \\
&& REM &
& 89.71 && 89.89 && 89.63 && 87.17 && 48.16 \\
&& Ours & 
& \textbf{88.79} && \textbf{88.36} && \textbf{88.25} && \textbf{84.84} && \textbf{23.12} \\
\midrule
\multirow{11}{1.2cm}{\centering CIFAR-100}
& \multirow{5}{0.5cm}{\centering $2/255$} & EM & \multirow{5}{0.5cm}{\centering 68.91} & 68.68 & \multirow{5}{0.5cm}{\centering 66.54} & 68.80 & \multirow{5}{0.5cm}{\centering 64.21} & 68.28 & \multirow{5}{0.5cm}{\centering 58.35} & 68.70 & \multirow{5}{0.5cm}{\centering 47.99} & 68.49 \\
&&  TAP &
& 68.40 && 67.93 && 67.25 && 67.09 && 66.91 \\
&& NTGA &
& 68.52 && 68.82 && 68.36 && 68.71 && 66.53 \\
&& REM &
& 68.90 && 68.29 && 61.42 && 51.99 && 16.60 \\
&& Ours & 
& \textbf{68.39} && \textbf{65.60} && \textbf{60.74} && \textbf{49.97} && \textbf{7.73}  \\
\cmidrule{2-13}
& \multirow{5}{0.5cm}{\centering $4/255$} & EM & \multirow{5}{0.5cm}{\centering 64.50} & 64.65 & \multirow{5}{0.5cm}{\centering 61.73} & 63.82 & \multirow{5}{0.5cm}{\centering 57.61} & 64.19 & \multirow{5}{0.5cm}{\centering 53.86} & 64.32 & \multirow{5}{0.5cm}{\centering 44.79} & 63.43 \\
&& TAP &
& 64.36 && 63.35 && 62.58 && 63.15 && 62.39 \\
&& NTGA &
& 63.48 && 63.59 && 63.64 && 62.83 && 62.44 \\
&& REM &
& 64.27 && 64.67 && 64.99 && 63.14 && 27.35 \\
&& Ours &
& \textbf{63.46} && \textbf{63.24} && \textbf{61.24} && \textbf{58.91} && \textbf{23.00} \\
\bottomrule
\end{tabular}
\vspace{-3mm}
\end{table}

Table~\ref{tab:percentage} shows that as the percentage of data protection decreases, the performance of the trained model increases. This suggests that the model can still learn from clean data, which is consistent with intuition. Furthermore, Table~\ref{tab:percentage} also shows that in all cases, the data protection ability of our method is stronger than that of other methods. This demonstrates that unlearnable noise generated by our method is more effective than other types of availability attacks even if mixed with clean data. This also hints that our methods can hide more information.  

\subsubsection{Transferablility}
\textbf{Different model architectures.}
Until now, we have only conducted adversarial training with ResNet-18, which is as same as the source model in the defensive noise generation.
We now evaluate the effectiveness of the unlearnable noise generated by our method under different adversarial training models.
Specifically, we conduct adversarial training with a perturbation radius of $4/255$ and five different types of models, including VGG-16, ResNet-18, ResNet-50, DenseNet-121, and wide ResNet-34-10, on data that is protected by noise generated via ResNet-18. The defensive perturbation radius $\rho_u$ of every type of defensive noise is set as $8/255$. 
Table~\ref{tab:diff_arch_at} presents the test accuracies of the trained models on CIFAR-10 and CIFAR-100.
\begin{table}[ht]
\centering
\caption{Test accuracy (\%) of different models on CIFAR-10 and CIFAR-100 datasets.}

\label{tab:diff_arch_at}
\scriptsize
\begin{tabular}{c c c c c c c c}
\toprule
\multirow{2}{1.2cm}{\centering Dataset} & \multirow{2}{0.8cm}{\centering Model} & \multirow{2}{0.8cm}{\centering Clean} & \multirow{2}{0.8cm}{\centering EM} & \multirow{2}{0.8cm}{\centering TAP} & \multirow{2}{0.8cm}{\centering NTGA} & \multicolumn{1}{c}{\centering REM} & \multicolumn{1}{c}{\centering Ours} \\
\cmidrule(lr){7-8}
& & & & & & $\rho_a = 4/255$ & $4/255$ \\
\midrule
\multirow{5}{1.2cm}{\centering CIFAR-10}
& VGG-16    & 87.51 & 87.24 & 86.27 & 86.65 & 65.23 & \textbf{37.78}    \\
& RN-18     & 89.51 & 88.62 & 88.02 & 88.96 & 48.16 & \textbf{23.12}    \\
& RN-50     & 89.79 & 89.66 & 88.45 & 88.79 & 40.65 & \textbf{19.30}    \\
& DN-121    & 83.27 & 81.77 & 81.72 & 80.73 & 82.38 & \textbf{72.42}    \\
& WRN-34-10 & 91.21 & 79.87 & 90.23 & 89.95 & 48.39 & \textbf{18.67}    \\
\midrule
\multirow{5}{1.2cm}{\centering CIFAR-100}
& VGG-16    & 57.46 & 56.94 & 55.24 & 55.81 & 58.07 & \textbf{55.05}    \\
& RN-18     & 64.50 & 63.43 & 62.39 & 62.44 & 27.35 & \textbf{23.00}    \\
& RN-50     & 66.93 & 66.43 & 64.44 & 64.91 & 26.03 & \textbf{21.47}    \\
& DN-121    & 53.73 & 53.52 & 52.93 & 52.40 & 56.63 & \textbf{52.25}    \\
& WRN-34-10 & 68.64 & 68.27 & 65.80 & 67.41 & 27.71 & \textbf{20.14}    \\
\bottomrule
\end{tabular}
\vspace{-3mm}
\end{table}

Table~\ref{tab:diff_arch_at} shows that our unlearnable noise generated from ResNet-18 can effectively protect data against various adversarially trained models and outperforms other methods by a significant margin. 


\textbf{Different Noise generators.}
Thus far, all the noise generated in our method has utilized ResNet-18 as a surrogate model. It is possible that the specific properties of ResNet-18 contribute to the effectiveness of our method. Therefore, to evaluate the generalization performance of our approach, we employ different surrogate models for generating unlearnable noise. Furthermore, since our ASR is not tied to any specific model architecture, it should remain effective regardless of the surrogate model employed. The adversarial training perturbation radius is configured to $4/255$, and the defensive perturbation radius $\rho_u$ for each type of defensive noise is set to $8/255$. We test four noise generators, including VGG-16, ResNet-18, ResNet-50, and DenseNet-121. Each noise variant is tested on five models, namely VGG-16, ResNet-18, ResNet-50, DenseNet-121, and WRN-34-10. We mainly focus on robust methods. 
Test accuracies on CIFAR-10 and CIFAR-100 are provided in Table~\ref{tab:trans_cifar10_at} and Table~\ref{tab:trans_cifar100_at}.

\begin{table}[ht]
\centering
\caption{Test accuracy (\%) of different models adverarially trained on unlearnable CIFAR-10 generated by different noise generators.}
\label{tab:trans_cifar10_at}
\scriptsize
\begin{tabular}{c c c c c c c c }
\toprule
{\centering Surrogate Model} & {\centering Method} & {\centering VGG-16} & {\centering ResNet-18} & {\centering ResNet-50} & {\centering DenseNet-121} & {\centering WRN-34-10} & {\centering Average} \\
\midrule
\multirow{3}{1.2cm}{\centering VGG-16}
& EM        & 87.75 & 89.21 & 90.19 & 83.58 & 90.83 & 88.31   \\
& REM       & 73.60 & 74.73 & 74.16 & 77.63 & 74.94 & 75.01   \\
& Ours        & \textbf{63.11} & \textbf{67.50} & \textbf{64.37} & \textbf{61.02} & \textbf{65.65} & \textbf{64.33}  \\
\midrule
\multirow{3}{1.2cm}{\centering ResNet-18}
& EM        & 87.24 & 88.62 & 89.66 & 81.77 & 79.87 & 85.43    \\
& REM       & 65.23 & 48.16 & 40.65 & 82.38 & 48.39 & 58.96    \\
& Ours        & \textbf{37.78} & \textbf{23.12} & \textbf{19.30} & \textbf{72.42} & \textbf{18.67} & \textbf{34.26}    \\
\midrule
\multirow{3}{1.2cm}{\centering ResNet-50}
& EM        & 87.57	& 89.17	& 89.83	& 82.64 & 90.68 & 87.98   \\
& REM       & 51.88 & 44.27 & 37.79 & 82.01 & 42.09 & 51.61   \\
& Ours        & \textbf{49.33} & \textbf{39.95} & \textbf{36.50} & \textbf{79.69} & \textbf{41.57} & \textbf{49.41} \\
\midrule
\multirow{3}{1.2cm}{\centering DenseNet-121}
& EM     & 87.59 & 84.51 & 85.57 & 82.76 & 85.68 & 85.22       \\
& REM    & 67.30 & 69.62 & 66.42 & 60.51 & 72.09 & 67.19       \\
& Ours     & \textbf{61.41} & \textbf{58.77} & \textbf{58.55} & \textbf{58.66} & \textbf{63.38} & \textbf{60.15}\\
\bottomrule
\end{tabular}
\vspace{-3mm}
\end{table}

\begin{table}[ht]
\centering
\caption{Test accuracy (\%) of different models adverarially trained on unlearnable CIFAR-100 generated by different noise generators.}
\label{tab:trans_cifar100_at}
\scriptsize
\begin{tabular}{c c c c c c c c }
\toprule
{\centering Surrogate Model} & {\centering Method} & {\centering VGG-16} & {\centering ResNet-18} & {\centering ResNet-50} & {\centering DenseNet-121} & {\centering WRN-34-10} & {\centering Average} \\
\midrule
\multirow{3}{1.2cm}{\centering VGG-16}
& EM        & 57.33 & 63.55 & 65.44 & 53.45 & 68.23 & 61.60     \\
& REM       & 41.13 & 52.00 & 51.77 & 48.92 & 56.05 & 49.97     \\
& Ours        & \textbf{36.67} & \textbf{45.82} &  \textbf{46.45} & \textbf{45.52} & \textbf{48.59} & \textbf{44.61}    \\
\midrule
\multirow{3}{1.2cm}{\centering ResNet-18}
& EM        & 56.94 & 63.43 & 66.43 & 53.52 & 68.27 & 61.72     \\
& REM       & 58.07 & 27.35 & 26.03 & 56.63 & 27.71 & 39.16     \\
& Ours        & \textbf{55.05} & \textbf{23.00} & \textbf{21.47} & \textbf{52.25} & \textbf{20.14} &  \textbf{34.38} \\
\midrule
\multirow{3}{1.2cm}{\centering ResNet-50}
& EM        & 56.82 & 64.19 & 66.93 & 54.51 & 68.56 & 62.20      \\
& REM       & 54.61 & 35.50 & 30.43 & 54.26 & 35.11 & 41.98     \\
& Ours        & \textbf{52.57} & \textbf{26.17} & \textbf{29.38} & \textbf{52.19} & \textbf{25.91} & \textbf{37.24} \\
\midrule
\multirow{3}{1.2cm}{\centering DenseNet-121}
& EM        & 57.39 & 63.73 & 66.37 & 54.62 & 68.43 & 62.11      \\
& REM       & 47.22 & 41.89 & 45.49 & 41.15 & 50.66 & 45.28     \\
& Ours        & \textbf{38.15} & \textbf{34.70} & \textbf{34.46} & \textbf{37.84} & \textbf{32.30} & \textbf{35.49}\\
\bottomrule
\end{tabular}
\vspace{-3mm}
\end{table}

can always outperform other availability attacks, whether they are robust or not.
As depicted in Table~\ref{tab:trans_cifar10_at} and Table~\ref{tab:trans_cifar100_at}, our method demonstrates exceptional generalizability. Regardless of the surrogate model's capabilities, our method consistently outperforms other availability attacks, both robust and non-robust. This indicates that our approach exhibits profound generalizability, and we have successfully proposed a generalizable method rather than a specific type of noise.

\subsubsection{Time Consumption}
Our method requires less time to train the robust noise generator than REM. Table~\ref{tab:time_cost} shows the time cost of three different methods. It should be noted that the time cost for CIFAR-10 and CIFAR-100 was tested using a V100, while the time cost for ImageNet-Subset was tested using $4$ V100s.
\begin{table}[ht]
    \centering
    \caption{Time consumption (h) of different methods on different datasets.}
    \begin{tabular}{c c c c c c }
    \toprule
         & EM & TAP & NTGA & REM & Ours \\
         \midrule
        CIFAR-10 & 0.4 & 0.5 & 5.2 & 22.6 & 5.6  \\
        CIFAR-100 & 0.4 & 0.5 & 5.2 & 22.6 & 5.6  \\
        ImageNet-Subset & 3.9 & 5.2 & 14.6 & 51.2 & 11.3  \\
        \bottomrule
    \end{tabular}
    \label{tab:time_cost}
    \vspace{-3mm}
\end{table} 

\section{Conclusion}

In this paper, we have systematically examined the limitations of existing availability attack methods, which aim to protect privacy by generating unlearnable noise in training data. We identified that the current optimization process of robust model-based availability attacks is suboptimal, bearing the risk of their protective effects being invalidated by adversarial training due to non-robust surrogate models. To address these issues, we proposed a two-stage optimization procedure for training robust surrogate models, which generates robust unlearnable noise resistant to adversarial training. Furthermore, we formalized the effect of unlearnable examples and modified the optimization objective for the robust surrogate model accordingly. Through comprehensive experiments, we demonstrated the superiority of our method in terms of effectiveness, speed, and generalization, thereby establishing a solid foundation for future research in this domain.

\textbf{Limitations.} 
The proposed method in this paper requires the introduction of an adversarial training process to generate robust unlearnable samples. This leads to a significant computational cost when extending the method to large-scale datasets, such as ImageNet. As a result, the scalability of the proposed approach may be limited, especially when dealing with massive datasets.
Additionally, the current method has not been optimized for scenarios involving partial protection of data. When adding unlearnable noise to only a portion of the data, the anti-learning effect is considerably weaker compared to the case where protective noise is added to the entire dataset. This gap highlights a valuable direction for future research, as it is essential to develop techniques that can effectively protect data privacy even when only a subset of the data is targeted for protection.

{
\small
\bibliographystyle{plain}
\bibliography{refs}
}

\end{document}